\documentstyle[aps,prb,preprint]{revtex} 
\begin{document}

\draft

\title{ Single Cooper pair tunneling induced by 
non-classical microwaves
       }
\author{
A. A. Odintsov$^{(1)}$ and A. Vourdas$^{(2)}$
}
\address{
$^{(1)}$Department of Applied Physics, Delft University of Technology,
2628 CJ Delft, The Netherlands \\
and  Nuclear Physics Institute, Moscow
State University, Moscow 119899 GSP, Russia \\
$^{(2)}$Department of Electrical Engineering and Electronics,
The University of Liverpool, Brownlow Hill, 
P.O. Box 147, Liverpool L69 3BX, England\\
} 
\date{\today}
\maketitle

\begin{abstract}
A mesoscopic Josephson junction
interacting with a mode of non-classical microwaves
with frequency $\omega$ is considered. 
Squeezing of the electromagnetic field
drastically affects the dynamics of Cooper tunneling.
In particular, Bloch steps
can be observed even when the microwaves are
in the squeezed vacuum state
with {\em zero} average amplitude of the field 
$\langle E(t) \rangle = 0$. The interval between these 
steps is double in size in comparison to the conventional
Bloch steps.

\end{abstract}

\pacs{PACS numbers: 73.40.Gk, 74.50.+r, 42.50.Dv}
%\twocolumn
%\narrowtext

There has been a lot of work \cite{1} on single charge 
tunneling in small junctions with current bias. 
The Bloch oscillations \cite{5} of the voltage with fundamental
frequency $f = I/2e$ has been observed \cite{bloch_obs}. 
This phenomenon is dual to the traditional AC Josephson effect. 
The same junctions biased with a current $I_{DC} + I_{AC} \sin(2\pi f t)$
produce a voltage that has a DC component when $I_{DC} = 2enf$ 
(so called Bloch steps which are dual to well known Shapiro steps). 
 
Non-classical electromagnetic fields (squeezed states) have been studied 
extensively in the last few years \cite{3}. They are sinusoidal signals in 
their average value of $<E>,<B>$ but they differ from each other in 
the quantum fluctuations $\Delta E, \Delta B$. At optical frequencies 
they are experimentally available in many laboratories, but they have 
also been  produced at microwave frequencies \cite{4}.
In recent work \cite{2} the AC Josephson effect has been studied
in the presence of non-classical electromagnetic fields.  
In this paper we study the effect of quantum electromagnetic fields
on a dual phenomenon - the Bloch oscillations.     

We consider a small Josephson junction in a resonant
cavity connected to a quantum source of microwaves.
A coupling of the junction to the microwaves can be achieved,
for example, by extending the electrodes of the junction
in the direction of the electric field (see inset of Fig.~1). 
The field polarizes a charge 
\begin{equation}
\hat{Q}_p = C d \hat{E}
\label{Q_p}
\end{equation}
at the electrodes of the junction (here $C$
is the total capacitance between the electrodes 
and $d$ is a coefficient of proportionality of the order
of the length $L$ of the electrodes).
The Hamiltonian of the Josephson junction has the form,
\begin{equation}
H_J = \frac {(\hat{Q}-\hat{Q}_p)^2}{2C} - E_J \cos \hat{\phi},
\label{H_J}
\end{equation}
where the charge $\hat{Q} = 2ne$ 
and the Josephson phase difference $\hat{\phi}$ across the 
junction are canonically conjugated operators,
$[\hat{Q}, \hat{\phi}] = -2ei$. 

We assume that the quantum source excites a single
mode of the microwaves with the frequency $\omega$.
In order to test a dynamics of the junction, it is
instructive to impose in addition a classical time-dependent 
component $E_{cl}(t)$ so that the total electric field is: 
\begin{equation}
\hat{E} = E_{cl} (t) + i E_\xi (a  - a^{\dagger}).
\label{E}
\end{equation}
Here $a^{\dagger}, a$ are creation 
and annihilation operators for the external non-classical 
electromagnetic field and
$E_\xi$ is a constant of the order of 
$(\hbar \omega / \epsilon_0 V)^{1/2}$,
$V$ being the volume of the resonator.
Note that in Eqs.~(\ref{Q_p}), (\ref{H_J}) we have neglected 
the dynamical redistribution of the charge on the junction
(characterized by relaxation time $\tau_{rel}$) and used 
adiabatic description of the Josephson tunneling. For this reason,
our model is valid at low frequencies 
$\omega \ll \min(\tau_{rel}^{-1}, \Delta/\hbar)$,
$\Delta$ being the superconducting gap.

The eigenvalues $E_{n}(\hat{Q}_p)$ and eigenfunctions 
$\Psi_{n,\hat{Q}_p}(\phi)$
of the Hamiltonian $H_J$ can be readily found in the basis of states,
in which the operator $\hat{Q}_p$ is diagonal. 
The energies $E_{n}(\hat{Q}_p)$ are $2e$-periodic functions 
of the polarization charge $\hat{Q}_p$ 
(which is analogous to a quasimomenta 
in the standard band theory).
We will assume that both 
classical ($E_{cl} (t)$) and quantum ($iE_\xi (a-a^{\dagger})$)
parts of the electromagnetic field change slowly enough in time
(see the condition (\ref{adiabcond}) below). For this reason,
we will consider only the lowest energy band $E_0 (\hat{Q}_p)$. 
The Hamiltonian of the system acquires a simple form,
\begin{equation}
H = E_0 (\hat{Q}_p)  + \hbar \omega (a^{\dagger}a + 1/2).
\label{H_ad}
\end{equation}

For strong Josephson coupling 
$E_J \gg E_C \equiv e^2/2C$ a dispersion relation 
for $E_0 (\hat{Q}_p)$ has the form~\cite{5}
\begin{equation}
E_0 (\hat{Q}_p) = \frac{\delta}{2} - \frac{w_0}{2} 
\cos \left( 2 \pi \frac{\hat{Q}_p}{2e} \right),
\label{largeE_J}
\end{equation}
where $w_0 = 32 \cdot 2^{-1/4} \pi^{-1/2} E_C^{1/4} E_J^{3/4} 
 \exp[-(8E_J/E_C)^{1/2}]$
is the widths of the band and the energy  
$\delta= (8E_C E_J)^{1/2}$ corresponds
to the gap between the lowest and the next energy bands.
 
% Discuss applicability of lowest band approximation
The lowest band approximation requires that 
both classical and non-classical parts of the polarization charge
$\hat{Q}_p$ change slowly on the time-scale determined by
the gap between the two lowest bands,
\begin{equation}
\langle (d\hat{Q}_p/dt)^2 \rangle^{1/2}/e 
\ll \delta/\hbar
\label{adiabcond}
\end{equation}
(the average is taken over the state the electromagnetic field, 
see below).
In what follows we will consider the simplest case when 
the Josephson junction perturbs the electromagnetic mode weakly, 
$w_0 \ll \hbar \omega$.
Clearly, this condition can be matched with the condition 
of adiabaticity (\ref{adiabcond}) at least in the limit of strong  
Josephson coupling (when $w_0 \ll \delta$).

In this limit the operator $\hat{V} = d E_0(\hat{Q}_p)/d \hat{Q}_p$ 
of the voltage across the junction 
is a harmonic function of $\hat{Q}_p$
(generally, it can be presented as a sum over the Fourier harmonics).
The voltage $V(t)$ is given by
the expectation value of the operator $\hat{V}(t)$ 
in the pure quantum state $| \Psi \rangle$ 
of the electromagnetic mode,
\begin{equation}
V(t) = V_0 \Im \left\{  
		e^{i q_{cl} (t)} 
		\langle \Psi |
		\exp[ \xi (a^{\dagger} e^{i \omega t} - a e^{-i \omega t}) ] 
		| \Psi \rangle 
		\right\},
\label{V(t)}
\end{equation}
where $V_0 = \pi w_0 / 2 e$, 
$q_{cl} (t) = \pi Q_p^{(cl)}(t)/e = \pi d E_{cl}(t) / (e/C)$ and
$\xi =  \pi d E_{\xi} / (e/C)$. 
We use the interaction representation
in which the time evolution of the operators is governed
by a free field Hamiltonian (second term in Eq.(\ref{H_ad})).

Specifically, we will consider squeezed states,
$| \Psi \rangle = \hat{D} \hat{S} | 0 \rangle$,
where squeezing ($\hat{S}$) and displacement ($\hat{D}$)
operators are given by~\cite{comsq} 
\begin{equation}
\hat{S} = \exp \left( \frac{r}{4} e^{-i \gamma} a^2 - 
		      \frac{r}{4} e^{i\gamma} {a^{\dagger}}^2
		\right),
\label{S}
\end{equation}
\begin{equation}
\hat{D} = \exp (\alpha a^{\dagger} - \alpha^{*} a).
\label{D}
\end{equation}
Using the transformation properties of $\hat{D}$ and $\hat{S}$ 
(namely, $\hat{D}^{-1} f(a, a^{\dagger}) \hat{D} = 
f(a+\alpha, a^{\dagger}+\alpha^{*})$
and analogous relation for $\hat{S}$ \cite{2,3})
together with the relation 
$\langle e^B \rangle = \exp(\langle B^2\rangle/2)$
for an arbitrary linear combination $B$
of Bose operators, we obtain from Eq.(\ref{V(t)}),
\[
V(t) = V_0 \sin \{ q_{cl}(t) + 2 \xi A \sin(\omega t - \chi) \}
\]
\begin{equation}
\times \exp \{ - (\xi^2/2) 
[ \cosh r + \sinh r \cos (2 \omega t - \gamma) ] \}
\label{Vgeneral}
\end{equation}
where $A$ and $\chi$ are the amplitude and the phase of the 
squeezed state, $\alpha = A e^{i \chi}$.

We consider now several applications of general formula
(\ref{Vgeneral}). First we assume that the classical component
of the field increases linearly in time, $q_{cl}(t) = \omega_{cl} t$,
which corresponds to a bias of the junction by the DC current
$I = 2e \omega_{cl}/ 2\pi$. 
One can see from Eq.~(\ref{Vgeneral}) that the voltage has a DC component 
$\bar{V}$ if $\omega_{cl} = n \omega$ with integer $n$.
This corresponds to a transfer of $n$ Cooper pairs through the
junction during the period $T= 2\pi/\omega$ of the electromagnetic
field oscillations. One can say that the transfer of Cooper pairs
is phase-locked by oscillations of the electromagnetic field.

Another way of interpretation is the following.
In the absence of electromagnetic field 
($\hat{Q}_p = 0$ in Eq.~(\ref{H_J})) the states of the Hamiltonian 
(\ref{H_J}) which differ by an integer number $k$ of the flux quanta 
($\hat{\phi} \to \hat{\phi} + 2 \pi k$) are degenerate.
In the presence of the classical field the part   
$Q_p^{cl} = 2e q_{cl}(t)/2\pi = e \omega_{cl} t/\pi$ 
of the Hamiltonian (\ref{H_J}) 
can be transformed into an additional term 
$- \hbar \omega_{cl} \hat{\phi}/2\pi$ 
of the Hamiltonian by a gauge transformation.
The states of the Hamiltonian are no more degenerate:
they form instead an equidistant
ladder with energy spacing $\hbar \omega_{cl}$.
The interaction with quantum-coherent component of electromagnetic field 
of frequency $\omega$ effectively restores the degeneracy of the states
with different number $k$ of the flux quanta
whenever the energy $n \hbar \omega$ matches with the 
level spacing $\hbar \omega_{cl}$. 
A disipationless tunneling of the Josephson phase 
between these states 
gives rise to a DC voltage $\bar{V}$ across the junction.

The voltage $\bar{V}$ can  be computed numerically by integrating 
Eq.~(\ref{Vgeneral}) over the period $T$.
The DC voltage is a function of the phases $\chi$ and $\gamma$.
The maximum $\bar{V}_{max}$ of the voltage 
with respect to these phases determines the maximum
amplitude of the Bloch steps.
For a pure coherent state ($r = 0$) we obtain,
\begin{equation}
\bar{V}_{max}^{(n)} = V_0 \exp(-\xi^2/2) J_n (2\xi A),
\label{Vcoherent}
\end{equation}
for $\omega_{cl} = n \omega$ and $\bar{V} = 0$ otherwise
(here $J_n(x)$ are the Bessel functions).
This expression differs from the result for a classical field
only by a constant factor $\exp(-\xi^2/2)$ which describes the effect 
of quantum fluctuations.

The situation is different for squeezed states ($r \neq 0$). 
Figure 1 shows the voltages $\bar{V}_{max}^{(n)}$
for the first four steps ($n =1,2,3,4$)
as functions of the coherent amplitude $A$ of a squeezed state.
One sees that the even steps 
$n=2,4$ exist even when the amplitude $A$ of the field is zero.
Indeed, for a squeezed vacuum ($A=0$)  Eq.~(\ref{Vgeneral}) gives
\begin{equation}
\bar{V}_{max}^{(2m)} = V_0 \exp \left(- \frac{\xi^2}{2} \cosh r \right) 
		   I_m \left( \frac{\xi^2}{2} \sinh r \right), 
\label{Vsqvac}
\end{equation}
for  $\omega_{cl} = 2m \omega$ and $\bar{V} = 0$ otherwise
(here $I_m(x)$ are the Bessel functions of imaginary argument).
This doubling of the interval between the steps 
is related to the fact that squeezed vacua 
are superpositions of  eigenstates with 
even number of photons. The amplitudes of the first four steps 
($n =2m =2,4,6,8$) as functions of the squeezing parameter $r$ 
are shown in Fig. 2.

Finally we consider the interaction of a classical AC field
$q_{cl}(t) = A_{cl} \sin (\Omega t)$
with quantum field in a squeezed vacuum state.
The voltage across the junction has a DC component
whenever the frequencies $\Omega$ and $2 \omega$ are commensurable,
(i.e. $\Omega/2\omega = p/q$, where $p$ and $q$ are coprime integers).
The amplitude of the Bloch step is given by:
\begin{equation}
\bar{V}_{max}^{(p,q)} = V_0 \exp \left(- \frac{\xi^2}{2} \cosh r \right) 
\sum_m J_{mq}(A_{cl}) I_{mp} \left( \frac{\xi^2}{2} \sinh r \right).
\label{VAC}
\end{equation}

We conclude with a comment on the relation between the present work and 
previous work on the electromagnetic environment 
(see Ch.~2 of Part 3 of Ref.~\cite{1} for a review). The results of 
previous papers were obtained for an environment in thermal equilibrium. 
In this work 
the external electromagnetic field is assumed to be in a particular 
quantum state; and the results depend on this state. 
The objective here is to use a carefully prepared quantum state 
of the electromagnetic field to influence the dynamics of the 
junction.

We would like to thank A.D. Zaikin for useful discussions.
The financial support of the European Community
through HCM Fellowship ERB-CHBI-CT94-1474 is gratefully acknowledged.
We also acknowledge the kind hospitality of ISI-Torino
(Italy) where part of this work was done.

\begin{figure}
\caption{Maximum amplitudes of the Bloch steps $n = 1, 2, 3, 4$
as functions of the amplitude $A$ of the squeezed state. 
We chose the parameters $\xi = 1$, $r = 1$. 
Inset: layout of the system.
\label{f1}}
\end{figure}

\begin{figure}
\caption{Maximum amplitudes of the Bloch steps $n = 2, 4, 6, 8$
as functions of the squeezing parameter for a
squeezed vacuum state. We chose $\xi = 1$.
\label{f2}}
\end{figure}


\begin{references}
\bibitem{comsq} 
Note that the definition of squeezed states here is different 
from that of Ref.~\cite{2}; both definitions are equivalent,
see e.g. part 1 of Ref.~\cite{3}.
\end{references}
\end{document}